\documentclass[12pt]{iopart}
\usepackage{amssymb}
\usepackage{bm}

\begin{document}

\title[Loop variables and gravitational Berry's quantum phase]{ Loop variables and gravitational
 Berry's quantum phase in the space-time of a rotating massive body}

\author{J. G. de Assis$^{1}$,  V. B. Bezerra$^{2}$ and C. Furtado$^{2}$}.

\address{
$^{1.}${ Departamento de Matem\'atica, Universidade Federal da
Para\'{\i}ba, 58051-970, Jo\~ao Pessoa, Pb, Brazil.}\\
$^{2.}${ Departamento de F\'{\i}sica, Universidade Federal da
Para\'{\i}ba, Caixa Postal 5008, 58051-970, Jo\~ao Pessoa, Pb, Brazil.}}

\begin{abstract}
In this paper we compute the holonomies along curves  
in the gravitational field of a slowly rotating massive body. We use our
results to study the gravitational analogue of Aharanov-Bohm effect 
in this space-time. We also investigate the behaviour of a scalar quantum 
particle in this space-time and determine Berry's quantum phase acquired
by this particle when transported along a closed curve surrounding the body.
\end{abstract}
 
\section{Introduction}

The loop space formalism for gauge theories was proposed in the beginning of the seventies
\cite{Mandelstam,Wilson}. In this formalism the fields depend on curves rather 
than on space-time points, and a gauge field is described by associating with each curve 
in space-time an element of the corresponding gauge group. The fundamental quantity that arises in this
approach, the non-integrable phase factor \cite{WuY}(loop variable) represents a gauge field 
more adequately than the field strength or the integral of the vector potential does. In this 
approach, the electromagnetism, for example, is a gauge-invariant manifestation of the 
non-integrable phase factor $exp({\frac{ie}{\hbar c}\oint_{c} A_{\mu}dx^{\mu}})$(loop variable), 
which gives the exact description of the theory \cite{WuY}, differently from the field strength 
and the integral of the potential, which underdescribes and overdescribes, respectively, the theory.

The extension of the loop space formalism  to the theory of gravity was firstly considered by 
Mandelstam\cite{Mandelstam} who established several equations involving loop variables, and also 
by Yang\cite{Yang}, Voronov and Makeenko \cite{Voronov} and  Bollini et al. \cite{Bollini} who 
computed the loop variables for the gravitational field corresponding to the Kerr metric. Other
investigations on loop variables in the context of gravitational fields includes the computation of this
mathematical object in the Schwarzschild-Droste geometry\cite{Ellis}, in Taub-NUT 
space-time\cite{Bini} and in the background space-time generated by a rotating black 
string\cite{Carvalho}. There are also studies on the connection between loop variables and 
gravitomagnetism\cite{Maartens} and clock effects\cite{Bini1}. 

Loop variables in the theory of gravity are matrices representing parallel transport along curves 
in a space-time with a given affine connection. They are connected with the holonomy linear 
transformation which contains importants topological informations. These mathematical objects 
contain information, for example, about how vectors change when parallel transported around a 
closed curve.

The appearance of topological phases  in the quantum  dynamics of a single
particle moving freely in multiply connected  space-times have been studied 
in a variety of physical systems. The prototype of this phase being the 
electromagnetic Aharonov-Bohm one\cite{AB}, which appears as a phase factor
in the wave function of an electron which moves around a magnetic flux line.
The gravitational analogue of this effect has also been investigated and 
discussed\cite{vil}.  

In the early eighties Berry discovered\cite{ber} that a slowly evolving(adiabatic) quantum
system retains informations of its evolution when returns to its original 
physical state. This information corresponds to what is termed Berry's phase.
The appearance of this phase has been generalized to the case of non-adiabatic
\cite{Ahan} evolution of a quantum system. In any case the phase depends only
on the geometrical nature of the pathway along which the system evolves. 
This phenomenon has been investigated in several areas of physics\cite{tom}.

Some works\cite{bro} concerning the investigation of Berry's phase in
the context of gravitation and cosmology were done in recent years. In 
particular, Cai and Papini\cite{cai} obtained a covariant generalization 
of the Berry's phase and applied this result to problems involving weak 
gravitational fields. Recently, Corichi and Pierri\cite{pie} studied the
behaviour of a quantum scalar particle in a class\, of\,\, stationary
space-times and investigated the phase acquired by the particle when 
transported along a closed path surrounding a rotating cosmic string.

This paper is organized as follows: in Section 2, we calculate the loop variables in
the space-time generated by a rotating massive body. In Section 3, we calculate Berry's
phase acquired by a scalar quantum particle in this gravitational field. Finally, in Section 4,
we present our remarks.

\section{Loop variables in the space-time of a rotating massive body}

To begin with, let us calculate the loop variables for different curves 
in the background space-time generated by a rotating massive body. In order 
to do this we shall consider the following expression for the loop variables\cite{VBB}
\begin{equation}
U_{AB}(C)=P\exp \left(- \int\limits_B^A\Gamma _\mu (x(\lambda ))\frac{dx^\mu }
{d\lambda }d\lambda \right) ,  \label{eq6}
\end{equation}
where $\Gamma _{\mu \textrm{ }}$ is the tetradic connection and $A$ and $B$
are the initial and final points of the curve. Then, associated with every
path $C$ from a point $A$ to a point $B$, we have a loop variable $U_{AB}$
given by eq.(\ref{eq6}) which, by construction, is a function of the curve $C$
as a geometrical object.

As the source of the gravitational field, we consider  an infinitely long, infinitely thin massive 
cylindrical shell which rotates slowly around its axis. The line element corresponding to this situation, 
in weak field approximation, is given by \cite{Frolov} 
\begin{equation}
ds^{2}=-(1-a(r)/2)dt^{2}+(1+a(r)/2)(dr^{2}+r^{2}d\phi ^{2}+dz^{2})+2b(r)dtd\phi,
\label{eq5}
\end{equation}
where $a(r)=-8m\Theta(r-r_{0})ln(r/r_{0}$  and  $b(r)=4j\left[ \frac{r^{2}}{r_{0}^{2}}\Theta
(r_{0}-r)+\Theta (r-r_{0})\right] $, with $\Theta(x)$ being the unit step function.

The parameter $m$ is the linear mass density and, $j=m\omega r_{0}$, is the linear angular momentum density, 
with $\omega $  being the angular velocity and  $r_{0}$ the radius of the cylindrical shell. This 
approximate solution is justified in a domain with $\left| a(r)\right|
=\left| -4\Phi \right| <<1$, where $\Phi $ is the Newtonian potential generated by the thin 
massive cylindrical shell and $b$ the angular momentum of the source, is such that $b^{2} \approx 0$.

The Riemann curvature tensor outside the shell is completely determined by the Newtonian potential, 
in the weak field approximation, if we neglect terms containing products of the 
form $ba(r)$, $a(r)^{2}$ and $b^{2}$. This means that in this approximation, the local effects of 
curvature associated with the rotation of the shell are absent outside it.

Now, let us compute the holonomy for general curves in the $xy$-plane. To do this 
let us first compute the tetradic connections. Defining the  1-forms

\begin{equation}
\begin{array}{ll}
\omega ^{0}= & (1-a/4)dt+bd\phi , \\ 
\omega ^{1}= & (1+a/4)dr, \\ 
\omega ^{2}= & (1+a/4)rd\phi , \\ 
\omega ^{3}= & (1+a/4)dz
\end{array}
\label{eq7}
\end{equation}
and using the Cartan structure equations, $d \omega^a= -\omega^{a}_{b} \wedge  \omega^{b}$, we obtain 
the tetradic connections which are given by
\begin{equation}
\begin{array}{ll}
\Gamma _{\mu 1}^{2}dx^{\mu }= & \left[ 1-2m(1-a/4)\right] d\phi =-\Gamma
_{\mu 2}^{1}dx^{\mu }, \\ 
\Gamma _{\mu 1}^{3}dx^{\mu }= & -2m/r(1-a/4)dz=-\Gamma _{\mu 3}^{1}dx^{\mu },
\\ 
\Gamma _{\mu 1}^{0}dx^{\mu }= & 2m/r(1-a/4)dt=\Gamma _{\mu 0}^{1}dx^{\mu },
\end{array}
\label{eq8}
\end{equation}
where $a(r)=-8m \ln (r/r_{0})$ and  $b(r)=4m\omega r_{0}$, outside the shell.

Following the usual procedure to calculate the holonomy\cite{VBB}, we shall first consider 
circles in the plane perpendicular to the cylinder with center 
at the origin, with fixed values of $t$ and $z$. So, in this case we have
\begin{equation}
\Gamma _{\mu }dx^{\mu }=\Gamma _{\phi }d\phi ,  \label{eq9}
\end{equation}
with  $\Gamma _{\phi }$ given by
\begin{equation}
\Gamma _{\phi }=\left( 
\begin{array}{cccc}
0 & 0 & 0 & 0 \\ 
0 & 0 & A_{\phi } & 0 \\ 
0 & -A_{\phi } & 0 & 0 \\ 
0 & 0 & 0 & 0
\end{array}
\right) ,  \label{eq10}
\end{equation}
where  $A_{\phi }=\left[ 1-2m(1-a/4)\right] $.

As  $\Gamma _{\phi }$ is independent of  $\phi ,$  the linear holonomy, $U_{L}$, for these circles is
\begin{eqnarray}
U_{L} &=&P\exp (\int\limits_{0}^{2\pi }\Gamma _{\phi }d\phi )=\exp (2\pi
\Gamma _{\phi })  \label{eq11} \\
&=&I+\frac{_{{}}\Gamma _{\phi }}{A_{\phi }}sen(2\pi A_{\phi })+\frac{\Gamma
_{\phi }^{2}}{A_{\phi }^{2}}[1-\cos (2\pi A_{\phi })],  \nonumber
\end{eqnarray}
where we used the fact that $(\Gamma _{\phi }^{{}})^{3}=-\Gamma _{\phi }$. Therefore in  matrix 
form the linear holonomy is given by
\begin{equation}
U_{L}(C)=\left( 
\begin{array}{cccc}
1 & 0 & 0 & 0 \\ 
0 & \cos 2\pi A_{\phi }^{{}} & sen2\pi A_{\phi } & 0 \\ 
0 & -sen2\pi A_{\phi } & \cos 2\pi A_{\phi }^{{}} & 0 \\ 
0 & 0 & 0 & 1
\end{array}
\right) =\exp (-2\pi iA_{\phi }J_{12}),  \label{eq12}
\end{equation}
with $J_{12}$ being the generator of the rotations about the z-axis.

Le us compute $U_{L}(C)$ for a translation in the $ z$-direction with $dt=dr=d\phi =0$. In this 
case we have 
\begin{equation}
\Gamma _{\mu }dx^{\mu }=\Gamma _{z}dz,  \label{eq13}
\end{equation}
where
\begin{equation}
\Gamma _{z}=\left( 
\begin{array}{cccc}
0 & 0 & 0 & 0 \\ 
0 & 0 & 0 & A_{z} \\ 
0 & 0 & 0 & 0 \\ 
0 & -A_{z} & 0 & 0%
\end{array}%
\right) ,  \label{eq14}
\end{equation}%
with  $A_{z}=2mr^{-1}(1-a/4)$. As  $\Gamma _{z}$ is independent of $z$, the loop variable 
corresponding to a segment that goes from $z_{1}$ to $
z_{2},U(C) $ is given simply by 
\begin{equation}
U_{z_{1}z_{2}}(C)=\exp \left( \int\limits_{z_{1}}^{z_{2}}\Gamma
_{z}dz\right) =\exp [-iA_{z}(z_{2}-z_{1})]J_{13},  \label{eq15}
\end{equation}
where $J_{13}$ is the generator of the rotations about $y$-axis.

For a translation in the time, we have $\Gamma _{\mu }dx^{\mu }=\Gamma _{t}dt, 
$ with  $\Gamma _{t}$ given by 
\begin{equation}
\Gamma _{t}=\left( 
\begin{array}{cccc}
0 & A_{t}\, & 0 & 0 \\ 
A_{t} & 0 & 0 & 0 \\ 
0 & 0 & 0 & 0 \\ 
0 & 0 & 0 & 0
\end{array}
\right) ,  \label{eq16}
\end{equation}
where  $A_{t}=2mr^{-1}(1-a/4)$. 
Therefore, for translation in the time between $t_1$ and $t_2$, we have the following
expression for the loop variable
\begin{equation}
U_{t_{1}t_{2}}(C)=\exp (-iA_{t}(t_{2}-t_{1}))J_{14},  \label{eq17}
\end{equation}
where $J_{14}$ is the generator of boosts in the $Oz$-direction. Note that the linear holonomy 
does not notice the angular momentum of the source.

Using these previous results given by eqs.(\ref{eq11}), (\ref{eq15}) and (\ref{eq17}), we can write
a general expression for the linear holonomy. In the general case, $U(C)$ reads

\begin{equation}
U(C)=Pexp(\frac{-i}{2} \int \Gamma^{ab}_\mu(x)J_{ab}dx^{\mu})\label{HOL},
\end{equation}
where $J_{ab}$ are the generators of the Lorentz group $SO(3,1)$ and $\Gamma_{\mu}^{ab}$ are the
appropriate tetradic connections.

Now, let us compute the translational holonomy which is associated with the jump in time when we 
go around the cylinder. Before going on, let us notice  that the line element given by 
(\ref{eq5}) can be put into the form 
\begin{equation}
ds^{2}=-[(1-a(r)/4)dt-bd\phi ]^{2}+(1+a(r)/2)(dr^{2}+r^{2}d\phi ^{2}+dz^{2}).
\label{eq18}
\end{equation}
Proceeding in analogy with the spinning cosmic string case\cite{VBB}, when we go around the 
cylinder 
along a circle from point $(t,x,y,z)$ to point 
$(t^{\prime },x^{\prime },y^{\prime },z^{\prime }),$  we have the following relations
\begin{equation}
\begin{array}{ll}
t^{\prime }= & (1-a/4)t-2\pi b, \\ 
x^{\prime }= & x\cos (2\pi A_{\phi })+ysen(2\pi A_{\phi }), \\ 
y^{\prime }= & -xsen(2\pi A_{\phi })+y\cos (2\pi A_{\phi }), \\ 
z^{\prime }= & z,
\end{array}
\label{eq19}
\end{equation}
where  $A_{\phi }=[1-2m(1-a/4)]$.

The transformations given by  eq.(\ref{eq19}) can be cast in the form of a homogeneous matrix 
multiplication by the following device: let $
M_{B}^{A}$ be a five dimensional matrix with $A$, $B$ running from $0$ to $4$. We take  
$M_{\mu }^{\nu }$ equal to the rotation matrix given eq.(\ref{eq12}), 
$M_{0}^{0}=-a/4$, $M_{4}^{0}=2\pi b$ and the other elements are equal to zero, so that 
\begin{equation}
\left( 
\begin{array}{l}
t^{\prime } \\ 
x^{\prime } \\ 
y^{\prime } \\ 
z^{\prime } \\ 
1%
\end{array}
\right) =(-iA_{\phi }J_{12}+i\frac{a}{4}M+ibM^{\prime })\left( 
\begin{array}{l}
t \\ 
x \\ 
y \\ 
z \\ 
1
\end{array}
\right)  \label{eq20}
\end{equation}
where 
\[
M=\left( 
\begin{array}{lllll}
0 & -i & 0 & 0 & 0 \\ 
0 & 0 & 0 & 0 & 0 \\ 
0 & 0 & 0 & 0 & 0 \\ 
0 & 0 & 0 & 0 & 0 \\ 
0 & 0 & 0 & 0 & 0
\end{array}
\right) { and  } \quad M^{\prime }=\left( 
\begin{array}{lllll}
0 & 0 & 0 & 0 & 0 \\ 
0 & 0 & 0 & 0 & 0 \\ 
0 & 0 & 0 & 0 & 0 \\ 
0 & 0 & 0 & 0 & i \\ 
0 & 0 & 0 & 0 & 0
\end{array}
\right) , 
\]
where the matrices $M$ and $M^{\prime}$ are such that  $[M,M^{\prime }]=0$, $M^{3}=-M$ e 
$M^{\prime 3}=-M$. Therefore, we can express the total holonomy for circle in the 
plane perpendicular to the cylinder as  
\begin{eqnarray}
U(C) &=&\exp (\frac{a\pi }{2}iM+2\pi ibM^{\prime })\exp (-2\pi iA_{\phi
}J_{12})  \label{eq21} \\
&=&\exp (-2\pi iA_{\phi }J_{12}+\frac{a\pi }{2}iM+2\pi ibM^{\prime }), 
\nonumber
\end{eqnarray}
where we are considering the five-dimensional representation of the generator of rotations
$J_{12}$.

Then, if one parallel transports a vector along a circle situated in the plane perpendicular to 
the massive body and around it, after this process it acquires a phase factor given 
by eq.(\ref{eq21}), which depends on the angular momentum. It is worth calling attention to the fact
that the angular momentum does not contribute to the curvature tensor in the weak field approximation. 
Therefore, a particle moving in the region where the Riemann curvature is different from zero and 
does not depend on the angular momentum exhibits a gravitational effect associated with the angular 
momentum. Then, we have an example of a non-local phenomena which corresponds to a gravitational 
analogue of the  Aharonov-Bohm effect\cite{vil}.
\section{Gravitational Berry's quantum phase in the space-time of a rotating massive body}
Now, let us calculate the Berry's quantum phase acquired by a scalar particle, in the
space-time of a rotating cylinder. The behavior of this particle is determined by the covariant 
Klein-Gordon equation

\begin{eqnarray}
  (\Box - M^{2}) \Psi=0, \label{eqi27}
\end{eqnarray}
where $\square =\frac{1}{\sqrt{-g}}\partial _{\mu }\left( \partial \sqrt{-g}%
g^{\mu \nu }\partial _{\nu }\right) $, $m$ is the mass of the particle and 
$\hbar=1$ and $c=1$ were chosen.

In order to do the calculation of the Berry's phase, let us write Klein-Gordon equation
in the background space-time under consideration, which is given by
 
\begin{equation}
\{(1+a)\frac{\partial ^{2}}{\partial t^{2}}-\frac{2b}{r^{2}}\frac{\partial
^{2}}{\partial \phi \partial t}-\frac{1}{r}\frac{\partial }{\partial r}(r
\frac{\partial }{\partial r})-\frac{1}{r^{2}}\frac{\partial ^{2}}{\partial
\phi ^{2}}-\frac{\partial ^{2}}{\partial z^{2}}+\frac{m^{2}(2+a)}{2}\}\Psi =0.
\label{eq23}
\end{equation}

Since the background space-time is time independent, we can look for solutions
of the form

\begin{equation}
\Psi=\exp(-iE_{n}t)\psi_{n}(r,\phi,z),\label{v1}
\end{equation}
where $E_{n}$ are the eigenvalues of energy.
Substituting eq.(\ref{v1}) into eq.(\ref{eq23}) we obtain an equation for $\psi(r,\phi,z)$. Then,
we consider the {\it ansatz } $\psi _{n}=\exp (i\xi \phi )\Phi _{n}$, and thus the equation
for $\Phi_{n}(r,\phi,z)$ reads
\begin{eqnarray}
\{\frac{\partial }{\partial r^{2}}+\frac{1}{r}\frac{\partial }{\partial r}+
\frac{1}{r^{2}}\frac{\partial ^{2}}{\partial \phi ^{2}}+\frac{\partial ^{2}}{
\partial z^{2}}-\frac{(2+a)}{2}m^{2}+ \nonumber \\ \frac{1}{r^{2}}(2i\xi -2biE_{n})\frac{\partial }{\partial \phi }+ \frac{1}{r^{2}}(-\xi ^{2} + 3bE_{n}\xi )+ (1+a)E_{n}^{2}\}\Phi _{n}(r,\phi,z)=0.\label{eq28}
\end{eqnarray}
\noindent
Now, let us take $\xi = bE_{n}$. This will imply that $\Phi _{n}$ obeys the equation
\begin{eqnarray}
\{\frac{\partial }{\partial r^{2}}+\frac{1}{r}\frac{\partial }{\partial r}+
\frac{1}{r^{2}}\frac{\partial ^{2}}{\partial \phi ^{2}}+\frac{\partial ^{2}}{
\partial z^{2}}-\frac{(2+a)}{2}m^{2}+(1+a)E_{n}^{2}\}\Phi _{n}=0,
\label{eq29}
\end{eqnarray}
\noindent
where we have neglected terms containing $b^2$. It is worth calling attention
to the fact that this choice for $\xi$ is such that eq.(\ref{eq29}) does not contain contribution 
coming from the angular momentum in the approximation considered. Thus we factorized out the 
contribution from the angular momentum of the source. In view of this result, we can use the Dirac 
phase method and write the solution
$\psi_{n}(r,\phi,z)$ as
\begin{equation}
\psi _{n}(r,\phi,z)=\exp \left( ibE_{n}\int\nolimits_{\varphi _{0}}^{\varphi
}d\phi \right) \Phi _{n}(r,\phi,z),  \label{eq31}
\end{equation}
with $\Phi_{n}(r,\phi,z)$  being the solution of the eq.(\ref{eq28}) 
which comes out from the Klein-Gordon equation in the background gravitational field given by  metric (\ref{eq5}), with 
 $b=0$. In this way we find the solution of a more complicated equation from the solution of a simpler one.

Now, let us calculate Berry's phase acquired by a scalar particle in the space-time under consideration. Taking into account that each eigenmode labeled by $n$ acquires a different geometric phase, and therefore we have a degeneracy of the energy eigenvalues, thus we have to follow the procedure done by Mostafazadeh\cite{Mosta} to appropriately treat this problem using the non-Abelian generalization of the Berry's phase, whose connection is given by
\begin{equation}
\mathcal{A}_{n}^{IJ}=<\Phi _{n}^{I}(\mathbf{x-R})\mid \nabla_{R}\Phi _{n}^{J}(\mathbf{x-R})>  \label{eq32}
\end{equation}
where $I$ and $J$ stand for possible degeneracy label. The vector $\mathbf{R}$ localizes the perfectly reflecting box which confines the quantum system, in relation to the center of the cylinder, and $\mathbf{x}$ localizes the particle relative to the center of the box.

The inner product in (\ref{eq32}) may be evaluated using the Dirac phase factor as follows:

\begin{eqnarray} \label{eq34}
<\psi _{n}^{I}({x_{i}-R_{i}})\mid \nabla _{R}\psi_{n}^{J}({x_{i}-R_{i}})> = \nonumber\\
=i\oint_{\Sigma }dS\{\psi _{n}^{\ast I}({x_{i}-R_{i}})[ibE_{n}\psi_{n}^{J}({x_{i}-R_{i}})+ \nabla_{R}\psi_{n}^{J}(x_{i}-R_{i})]\}  \nonumber\\ 
 =bE_{n}\delta _{IJ}dR^{2}  \nonumber
\end{eqnarray}
where $S=rdrd\phi dz$ and $R^{2}$ is the polar angle in the box.

Berry's phase can be obtained from the previous result. It is given by 
\begin{equation}
\gamma _{n}(C)= 2\pi bE_{n}
\label{eq35}
\end{equation}%
where $C$ is a curve which involves the cylinder.
\section{Final remarks}
We have shown by explicit calculation, for a metric corresponding to a slowly  rotating
massive body that the phase shift acquired by a particle, when
parallel transported along a given curve $C$ in this background gravitational field
is given by an element of the Lorentz group $SO(3,1)$ times a factor which depends on
the angular momentum of the source. From the results concerning the holonomy for curves in the plane perpendicular to the
cylinder, it comes out a phase which depends on the angular momentum of the source and this
corresponds to the gravitational analogue\cite{vil} of the Aharonov-Bohm effect.

Berry's phase for a scalar particle induced by the gravitational field under consideration 
also depends on the angular momentum of the source and was obtained by the non-Abelian 
generalization\cite{Mosta} of this phase, in order to take into account the degeneracy
of the energy levels of the system. 
\ack{This work was partially supported by CNPq and CAPES (PROCAD).
}\\

\end{document}